\documentclass[10pt, twocolumn]{article}

\usepackage{amsmath, amsthm, amssymb, amsfonts, latexsym}
\usepackage{graphicx}
\DeclareGraphicsExtensions{.jpeg}

\title{Topological De-Noising: Strengthening the Topological Signal}
\author{Jennifer Kloke and Gunnar Carlsson\thanks{The first author was supported by an NSF Graduate Fellowship.  The second author was supported by DARPA HR-0011-05-1-0007 and NSF grants DMS-0354543 and DMS-0406992.}}

\begin{document}
\maketitle

\begin{abstract}
Topological methods, including persistent homology, are powerful tools for analysis of high-dimensional data sets but these methods rely almost exclusively on thresholding techniques.  In noisy data sets, thresholding does not always allow for the recovery of topological information.  We present an easy to implement, computationally efficient pre-processing algorithm to prepare noisy point cloud data sets for topological data analysis.  The topological de-noising algorithm allows for the recovery of topological information that is inaccessible by thresholding methods.  We apply the algorithm to synthetically-generated noisy data sets and show the recovery of topological information which is impossible to obtain by thresholding.  We also apply the algorithm to natural image data in $\mathbb{R}^8$ and show a very clean recovery of topological information previously only available with large amounts of thresholding.  Finally, we discuss future directions for improving this algorithm using zig-zag persistence methods.
\end{abstract}


%



\section{Introduction}
An important challenge of data mining is the need to generate tools to simplify data down to important features or relationships.  Topology, the mathematics that arises in an attempt to describe global features of a space via local data, can provide new insights and tools for finding and quantifying relationships particularly on data that is not practical to understand using standard linear methods.  The development of algorithms that compute topological invariants from realistic data to identify important internal relationships is quite recent  \cite{topologyData}.  One of the most well-developed of these algorithms, persistent homology, provides qualitative information, such as connectedness and the presence of holes, in point cloud data (see \cite{computingPersistent} for more details) and has been shown to be a powerful tool for high-dimensional data analysis \cite{localbehavior}.  (See section~\ref{sec:topintro} for an introduction to topological analysis.)



\subsection{The Need for Topological De-Noising}\label{section:bumpintro}

While topological methods are very useful for data analysis, outliers create problems for any topological data analysis method that is applied to a whole data set.  Thresholding, the technique that filters the data based on some parameter, such as density, and returns a percentage of the points within a range of that parameter, such as only the top $10\%$ densest points, has been useful in limiting the effects of noise and, currently, topological data analysis relies mostly on thresholding techniques.  Thresholding, however, does not always allow for the recovery of the correct topological structure; this is especially true in instances were the level of noise is quite high.   

We have encountered a number of real data sets generated from experimental scientists from a variety of fields that, for external reasons, we believe contain non-trivial topological information.
The failure of thresholding methods, however, to limit the effects of noise in these instances has rendered topological methods ineffective.  A simple example where density thresholding does not allow persistent homology to recover the correct topological structure can be seen in a noisy sampling of the unit circle in $\mathbb{R}^2$.  Section~\ref{sec:noisycircle} is a detailed discussion of this example.

Standard de-noising or regularization methods, such as Tikhonov regularization and the conjugate gradient method, often quickly become computationally inefficient as the dimension of the data increases.  Principal curves \cite{principalcurves}, the generative topographical mapping \cite{gtm}, and graph Laplacian methods for manifold denoising \cite{graphlaplacian} all address the problem of identifying a noisily sampled submanifold $M$ in $\mathbb{R}^n$.  The first two methods require the user to know the intrinsic dimension of the submanifold, which can be difficult to estimate in the presence of high-dimensional noise.  Graph Laplacian methods are limited to finding low-dimensional submanifolds in $\mathbb{R}^n$. 

\subsection{Summary of Contributions} 
In section~\ref{section:alg} we present a computationally tractable technique that can take noisy data as its input, possibly in high dimensions, and return data that has less noise but still retains the underlying topological information from the original data set regardless of the dimension of the underlying manifold.  This is followed by a discussion of the intuition associated with this algorithm.  In section~\ref{section:3}  we apply the algorithm to synthetically-generated noisy data sets and show the recovery of topological information which is impossible to obtain by thresholding.  Section~\ref{section:mumford} contains the results of applying this algorithm to natural image data in $\mathbb{R}^8$ and shows a very clean recovery of topological information previously only available with large amounts of thresholding.  Finally, we discuss future directions for improving this algorithm using zig-zag persistence methods.

\subsection{Introduction to Topological Data Analysis}\label{sec:topintro} 

Persistent homology provides information about a point cloud data set by computing the homology of a sequence of nested spaces built on the data set.  Generally, we say that the homology of a space captures the connectivity, number of loops, and higher dimensional voids in the space.  The homology of a space can be represented as Betti numbers, written $\{\beta_n\}$, which gives a count of the $n$-dimensional voids in the space.  More specifically, $\beta_0$ measures the number of components in the simplicial complex; $\beta_1$ counts the number of  \textit{tunnels}, loops that cannot be deformed to a point in the complex; and $\beta_2$ counts the number of \textit{voids}, enclosed spaces that look like a room.  

For example, if $X$ is the unit circle in the plane, its non-zero Betti numbers are $\beta_0 = \beta_1 = 1$ since $X$ is connected and has a non-trivial loop that cannot be deformed to a point \textit{while staying in the space.} If $X$ is the figure eight, its non-zero Betti number are $\beta_0 = \beta_1 = 2$ since it has two non-trivial loops that cannot be deformed to a point or to each other.  If $X$ is the space resulting from removing an open unit ball from $\mathbb{R}^3$, its non-zero Betti numbers are $\beta_0 = \beta_2 = 1$ because $X$ is connected and has a ``two-dimensional'' non-trivial loop. If we had removed $k$ disjoint balls from $\mathbb{R}^3$, $\beta_2$ would have been $k$.

To compute homology from point cloud data $D$, the algorithm of persistent homology builds a sequence of nested simplicial complexes with the points in $D$ being the vertices of the complexes (see  \cite{topologyData} for more details).  Specifically, a \textit{simplicial complex} $K$ is a space built of vertices, edges, triangles, and their higher-dimensional counterparts (called simplices) where any face of a simplex in $K$ is also in $K$ and the intersection of any two simplices is the face of both of those simplices.  Nested simplicial complexes are built from a point cloud data as follows.  Let $\{\epsilon_n\}$ be an increasing sequence of positive real numbers.  Define the simplicial complex $K_n$ for each $\epsilon_n$ so that the vertex set is the data $D$ and two points $p, q \in D$ are connected by an edge in $K_n$ if the distance between $p$ and $q$ is less than $\epsilon_n$.  Moreover, any subset of $m$ points in $D$ form an $m-$simplex in $K_n$ if all the points in the subset are within $\epsilon_n$ distance.  This simplicial complex is known as a Vietoris-Rips complex.  It is not hard to see that that if $\epsilon_n < \epsilon_m$ then $K_n \subseteq K_m$.

\begin{figure}[h]
\centering
\includegraphics[width=.25\textwidth]{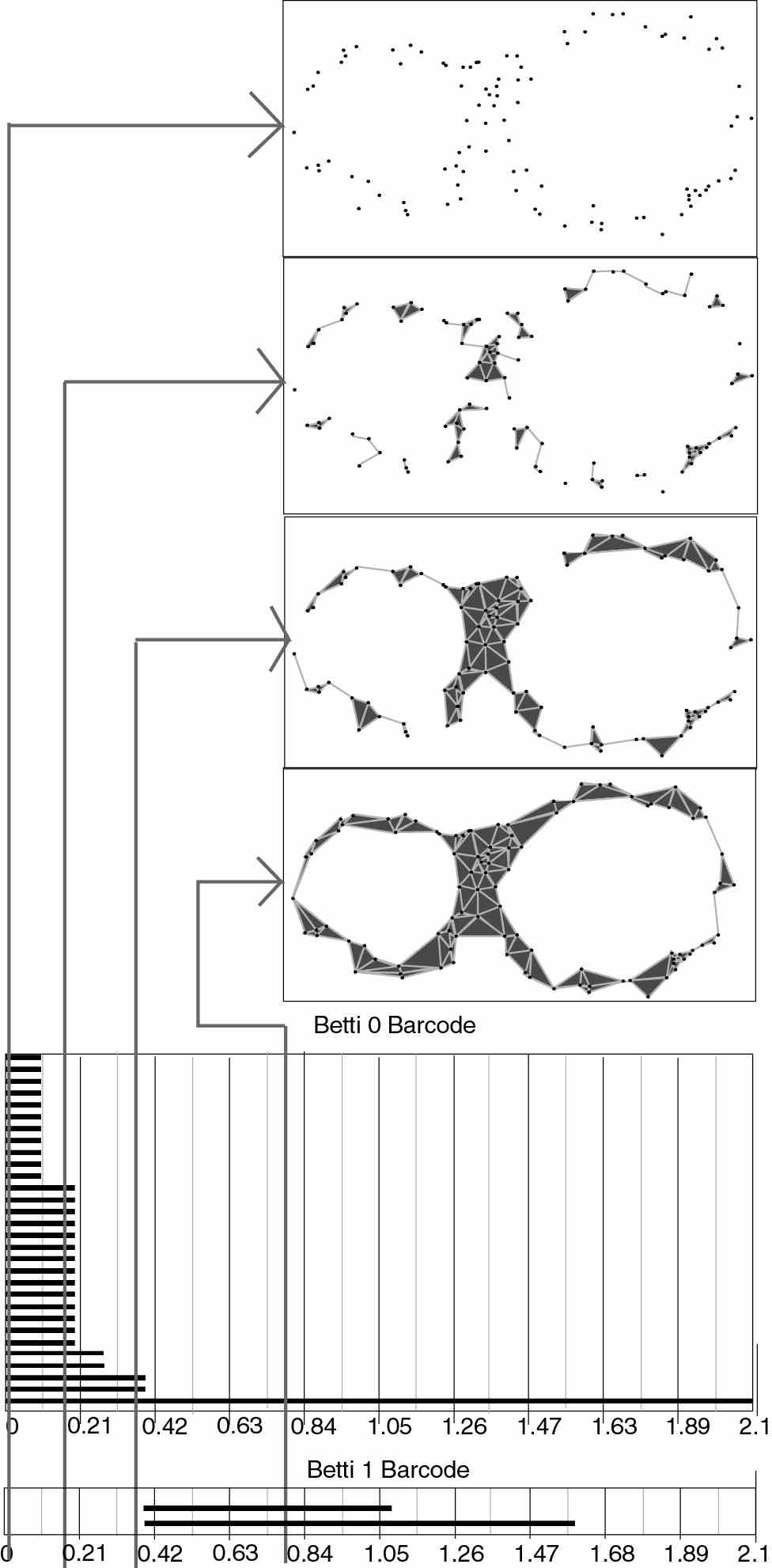}
\caption{Reading persistence barcodes.  The point cloud data set here is a sampling of a figure eight.  A nested sequence of simplicial complexes is built. The $x$-axis on corresponds to the maximal distance allowed for including simplices these complexes.  The number of lines above an $x-$value is the Betti number of the corresponding simplicial complex.  Horizontal lines in the Betti $n$ barcode indicate $n-$dimensional voids that persist from one complex to the next in the sequence.  A long line in a Betti $n$ barcode is evidence that there is an $n-$dimensional void in the data.}\label{fig:Barcode}
\end{figure}

The homology is computed for each complex and, through a special property of topology, we can determine which $n$-dimensional loops persist from one simplicial complex to the next complex in the sequence.  We use this information to generate barcodes to display the homology of this sequence of complexes.  For each dimension, the number of Betti numbers for each complex is plotted and if an $n$-dimensional loop persists from one complex to the next, a line in the barcode connects the Betti numbers for those complexes.  Figure ~\ref{fig:Barcode} is an example of a sequence of simplicial complexes built from a point cloud data set sampled from a figure eight as well as the corresponding 0- and 1-dimensional barcode associated to that sequence.

If a line, say the Betti 1 barcode, persists through many of these nested simplicial complexes, then this is evidence that a 1-dimensional hole is present in the data.  For instance, in figure~\ref{fig:Barcode}, the long bar for the Betti 0 barcode indicates that the data can be considered as a single connected component (or, in other words, a single cluster).  Likewise, the two long lines for the Betti 1 barcode indicate that there are two 1-dimensional holes in the data.

\subsection{Example where Thresholding Fails}\label{sec:noisycircle}

As mentioned in section~\ref{section:bumpintro}, thresholding fails to recover the correct topological structure from a noisy sampling of the unit circle in $\mathbb{R}^2$.  We can create such a data set by sampling the probability density function generated by convolving the delta function that has support being the unit circle in $\mathbb{R}^2$ with the spherically-symmetric 2-dimensional Gaussian function of standard deviation $\sigma$.  More specifically, we sample the probability distribution generated by normalizing the following function:
\[
p_{\sigma}(\vec{x})=\iint\limits_{\mathbb{R}^2} \, \delta_C(\vec{y}) e^{\left(\frac{-||\vec{y}-\vec{x}||^2}{2\sigma^2}\right)}\,d\vec{y}
\]
where $\delta_C(\vec{x})$ is the delta function that has support being the unit circle in $\mathbb{R}^2$.

\textit{Note:} By sampling the probability density function, we mean that we create a set of $N$ points via the following method:  randomly select a point $x$ within the support of the function $p$ and then randomly select a number between 0 and 1.  If $p(x)$ is greater than the selected number, $x$ is included in the set; otherwise, $x$ is disregarded.  Continue until the set contains $N$ points.

\setlength\fboxsep{0pt}
\setlength\fboxrule{0.5pt}
\begin{figure}[htbp]
\centering
\fbox{\includegraphics[width=.25\textwidth]{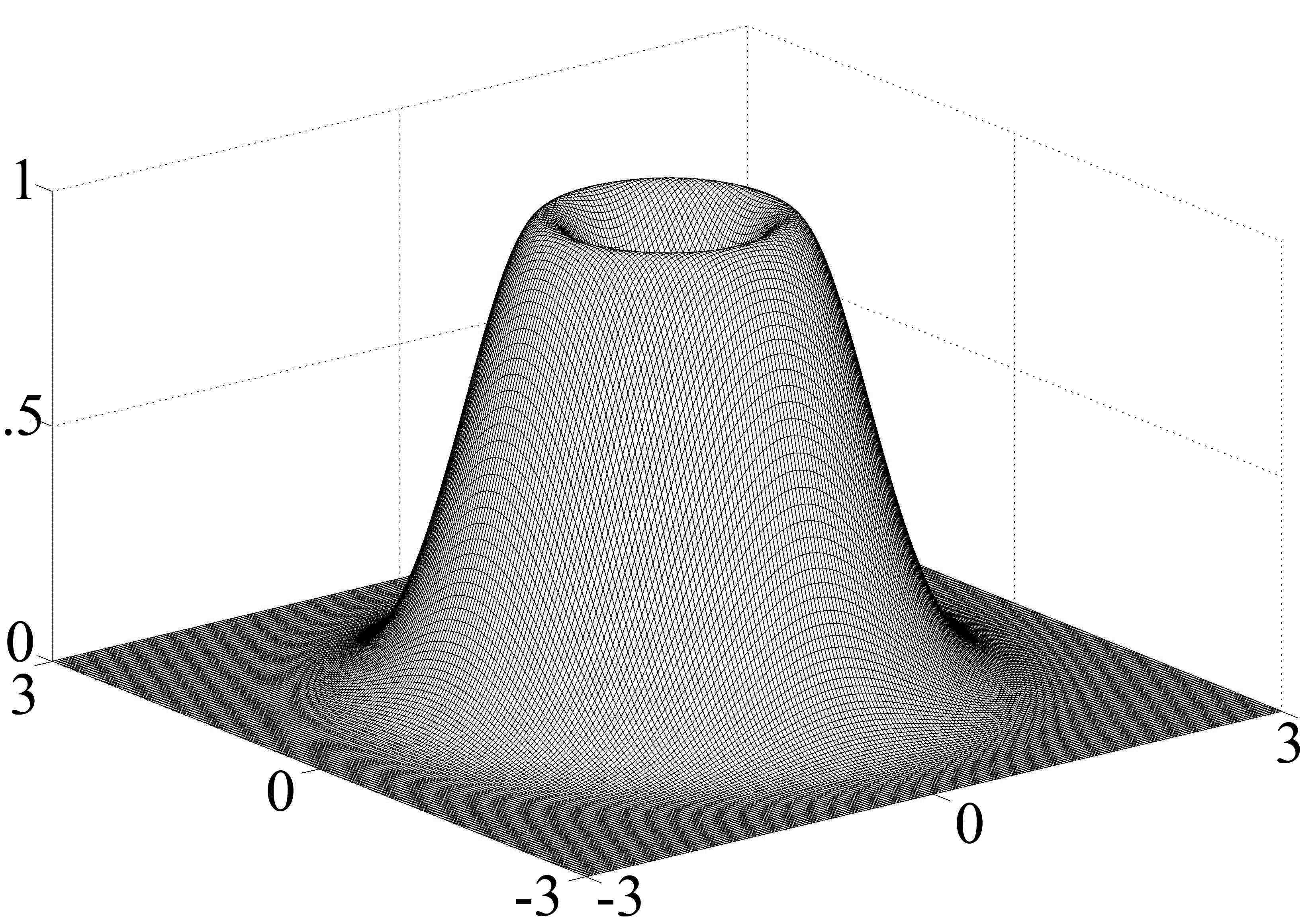}}
\caption{Probability density function for the unit circle in $\mathbb{R}^2$ with a spherically-symmetric Gaussian noise of standard deviation $\sigma= 0.5$}\label{fig:BumpWithCrater}
\end{figure}

\setlength\fboxsep{0pt}
\setlength\fboxrule{0.5pt}
\begin{figure}[htbp]
\centering
\fbox{\includegraphics[width=.25\textwidth]{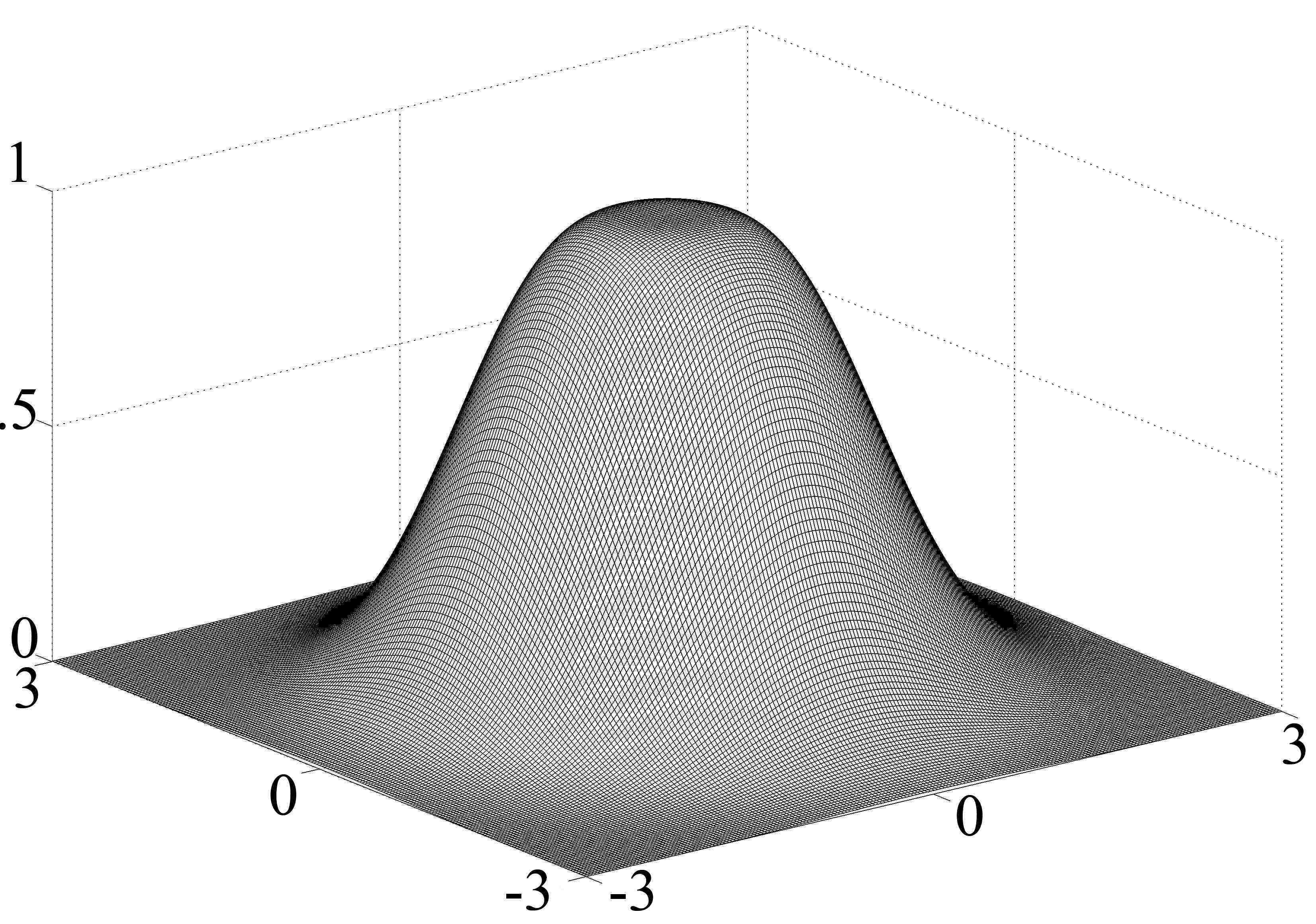}}
\caption{Probability density function for the unit circle in $\mathbb{R}^2$ with a spherically-symmetric Gaussian noise of standard deviation $\sigma= 0.7$}\label{fig:Bump_7}
\end{figure}

\begin{figure}[htbp]
\centering
\fbox{\includegraphics[width=.25\textwidth]{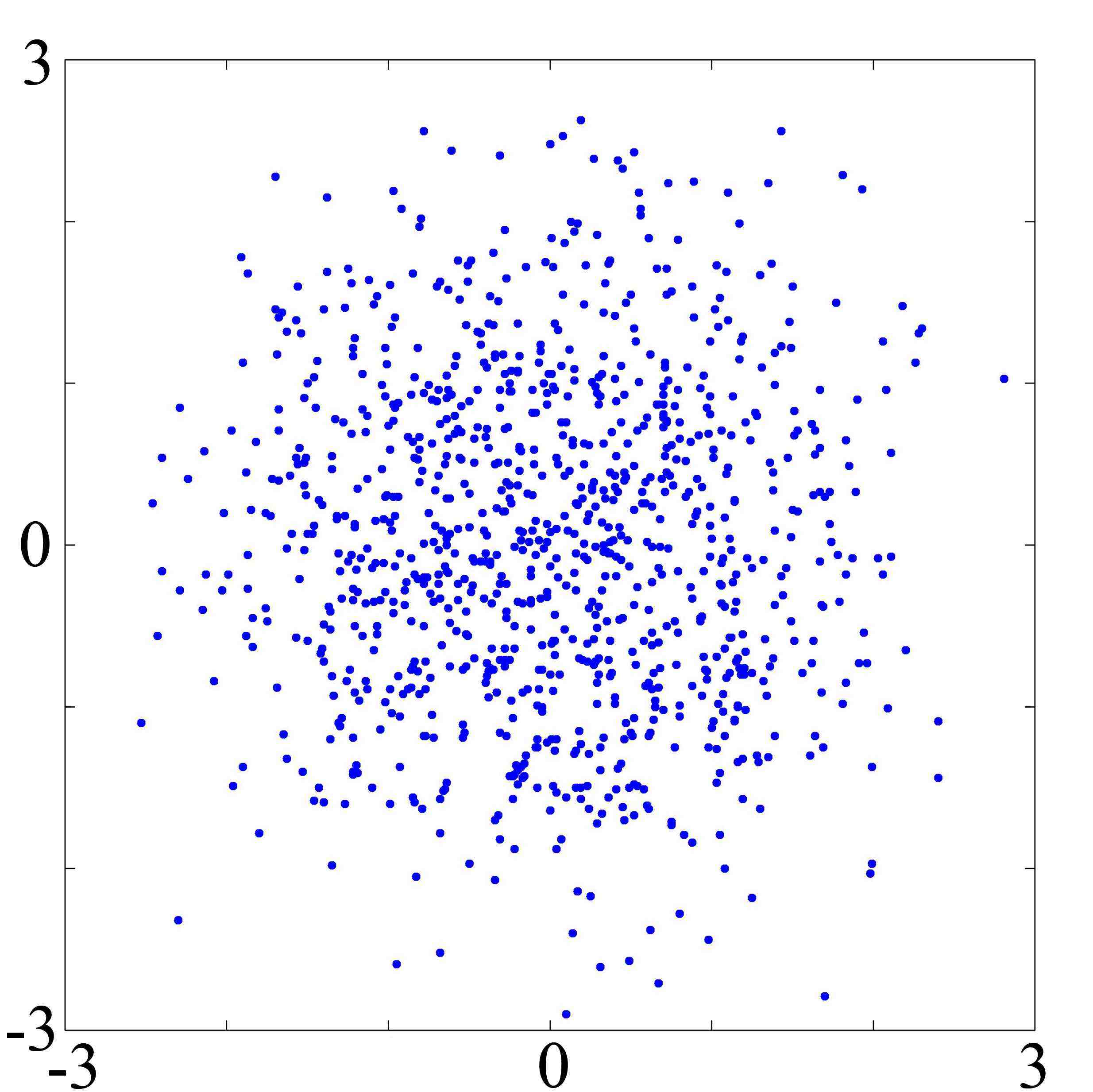}}
\caption{A noisy sampling of the unit circle in $\mathbb{R}^2$.  This point-cloud data set $K$ contains 1000 points sampled from the distribution in Figure 1}\label{fig:P7}
\end{figure}

\begin{figure}[h]
\centering

\fbox{\includegraphics[width=.25\textwidth]{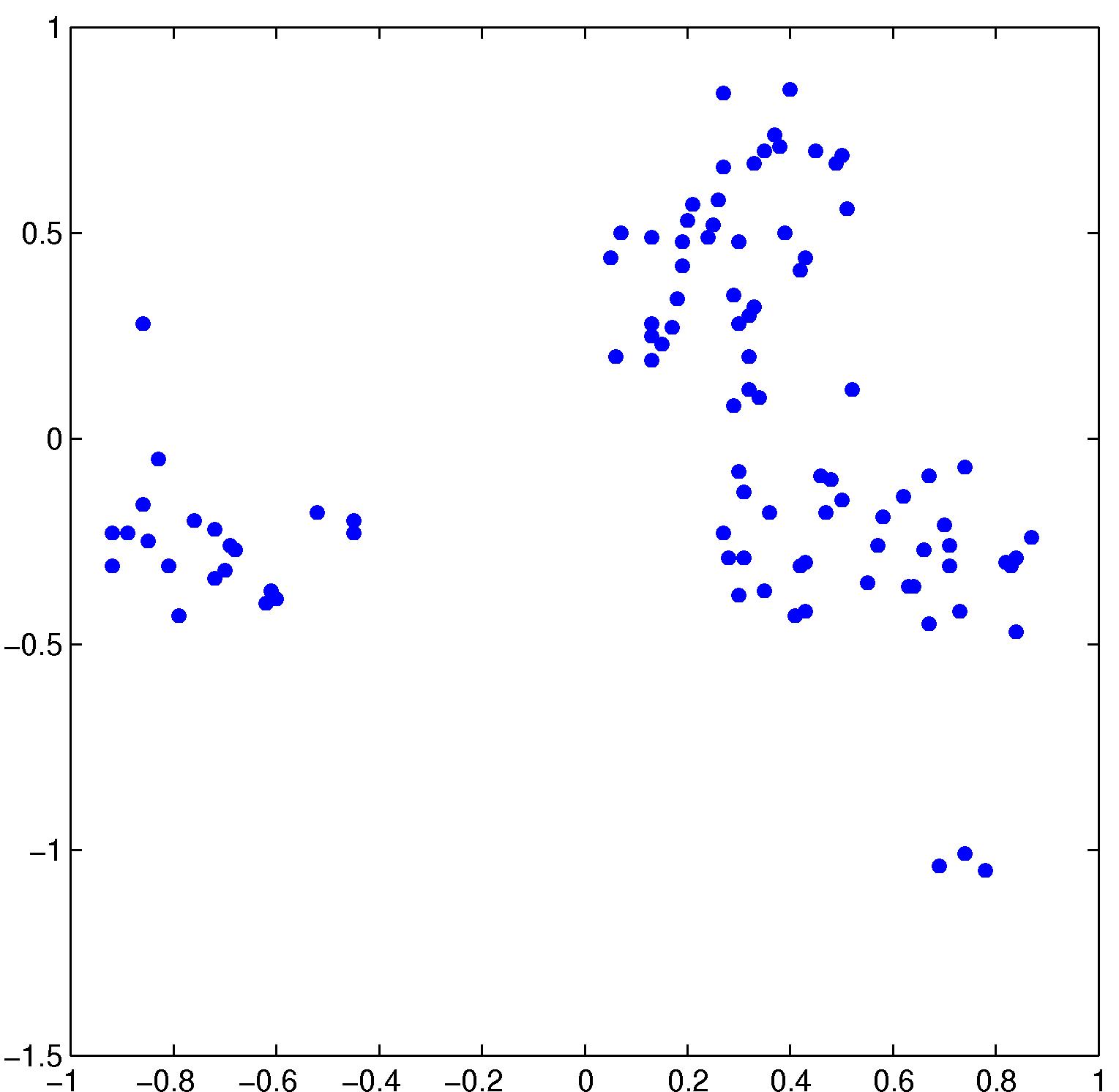}}
\caption{An example of thresholding applied to the data set $K$ from figure~\ref{fig:P7}.  Ten percent densest points in $K$ using the 30th nearest neighbor method to estimate density}\label{fig:thresh1}
\end{figure}

\begin{figure}[h]
\centering
\fbox{\includegraphics[width=.25\textwidth]{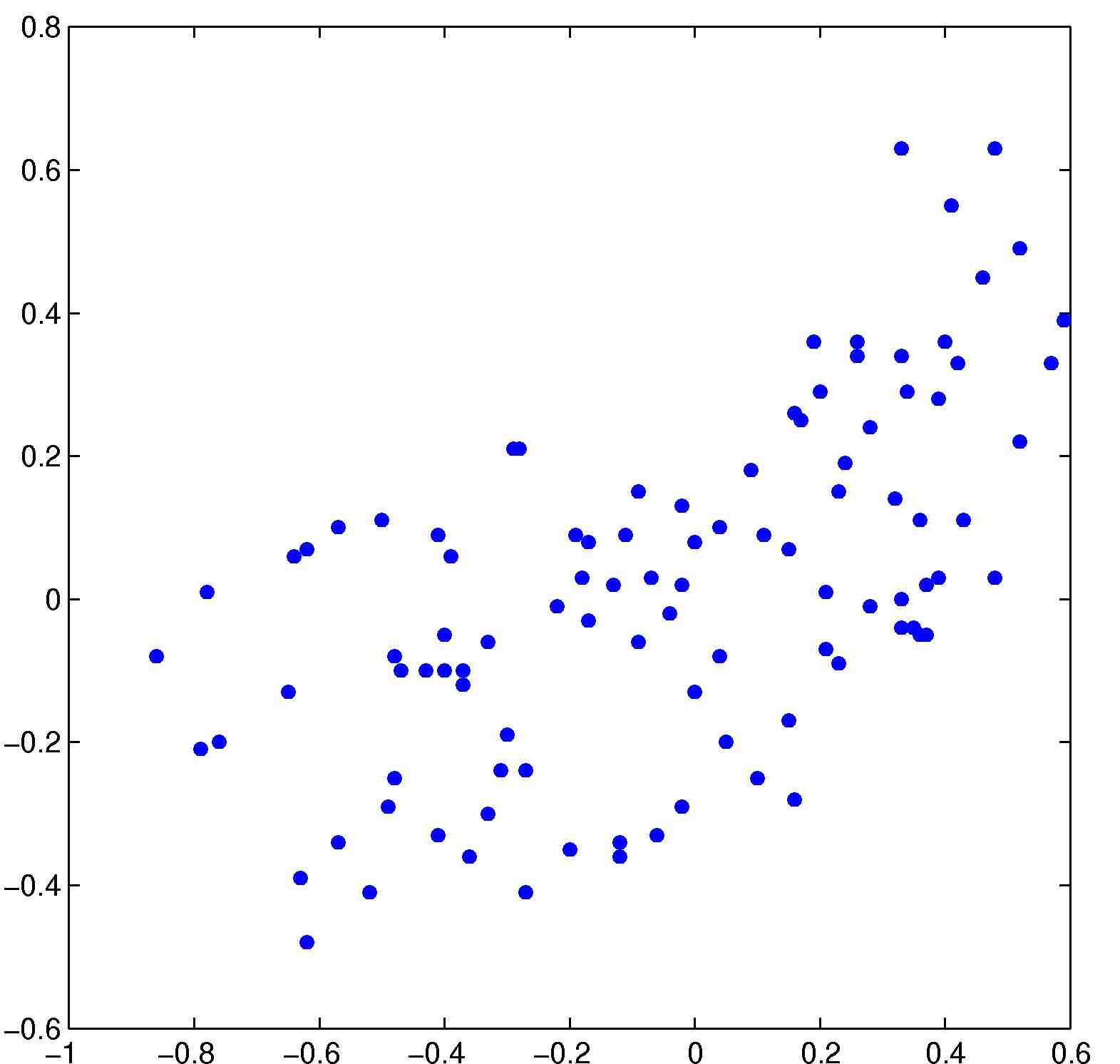}}
\caption{An example of thresholding applied to the data set $K$ from figure~\ref{fig:P7}.  Ten percent densest points in $K$ using the 75th nearest neighbor method to estimate density}\label{fig:thresh2}
\end{figure}

When the standard deviation $\sigma$ on the Gaussian function in $p(\vec{x})$ is small, the resulting density function has a significant crater over the origin (see figure~\ref{fig:BumpWithCrater}).  Persistent homology with thresholding recovers the generating circle from data sampled from such a density function.  If, on the other hand, the density function has a very shallow crater or none at all (such as the function in figure~\ref{fig:Bump_7}), thresholding does not allow for the recovery of the generating circle.

In the following example we sample from a density function $p_{0.7}(\vec{x})$ generated with a Gaussian function that has standard deviation $\sigma=0.7$ (see Figure~\ref{fig:Bump_7}.)  The point-cloud data set $K$ in figure~\ref{fig:P7} is a sampling of 1000 points from this probability density function.

Topological analysis of the data set $K$, with any level of density thresholding, fails to recover a persistent first Betti number of one (refer to section~\ref{sec:topintro} for explanation of Betti numbers and barcodes).  For thresholding in this paper, we use the nearest neighbor estimation of the density of a point cloud data set $X$.  The nearest neighbor estimator at $x \in D$ is defined to be inversely proportional to the distance from the point $x$ to the $k$-th nearest neighbor of $x$ (see \cite{silverman} for more details).  Figures~\ref{fig:thresh1} and~\ref{fig:thresh2} are two examples of density thresholding on the data set $K$ in figure~\ref{fig:P7}.  The data set in figure~\ref{fig:thresh1} contains the $10\%$ densest points of $K$ when the 30th nearest neighbor density estimator is used.  The data set in figure~\ref{fig:thresh2} contains the $10\%$ densest points of $K$ when the 75th nearest neighbor density estimator is used.

\begin{figure}[h]
\centering
\includegraphics[width=.45\textwidth]{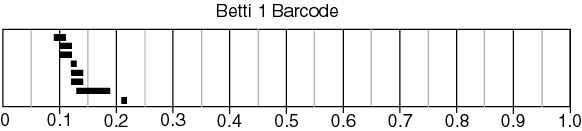}
\caption{Persistence barcode for Betti 1 of the $10\%$ densest points of $K$ using $30th$ nearest neighbor density estimation}\label{fig:barcodethresh}
\end{figure}

\begin{figure}[h]
\centering
\includegraphics[width=.45\textwidth]{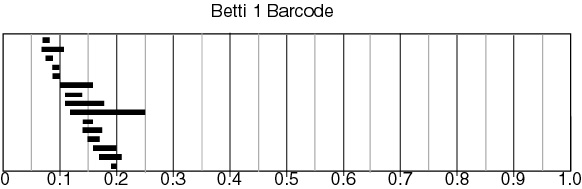}
\caption{Persistence barcode for Betti 1 of the $10\%$ densest points of $K$ using $75th$ nearest neighbor density estimation}\label{fig:barcodethresh2}
\end{figure}

Figures~\ref{fig:barcodethresh} and~\ref{fig:barcodethresh2} are two examples of persistent homology barcodes for Betti 1 of the data sets in figures~\ref{fig:thresh1} and~\ref{fig:thresh2}, respectively.  These barcodes were generated using Vietoris-Rips complexes built on the respective data sets.  One can see that both thresholded data sets fail to recover a persistent first Betti number of one (which indicates the presence of a circle in the data).  Again, density thresholding of the data set $K$ with a higher percentage of the data set or with different values of $k$ used to compute the $k$-th nearest neighbor density estimates fail to recover the correct topology from this data set.  Thresholding also fails to recover the topology from data sets created from similar probability density functions as in figure~\ref{fig:BumpWithCrater} with smaller standard deviations as low as $\sigma= 0.5$. 

\section{Topological De-Noising \\
Algorithm}\label{section:alg}
Let $\mathbb{X} \subseteq \mathbb{R}^d$ and let $D$ be a finite set of points sampled from a probability density function on $\mathbb{R}^d$ based around $\mathbb{X}$.  In other words, $D$ is a noisy sampling of the space $\mathbb{X}$.  Let $\sigma$ be an estimate on the standard deviation of the probability density function used to generate $D$.

Let $S_0$ be a subset of points randomly selected from $D$ (for example, $S_0$ might have $10\%$ of the points in $D$).  We will recursively define new data sets $S_n$ as follows.

For each $n \in \mathbb{N}$, define $F_n(x)$ to be the weighted difference of the average of spherically-symmetric Gaussian functions of standard deviation $\sigma$ centered at points in $D$ and $S_n$,
\[
F_n(x)  =   \frac{1}{|D|}\sum_{p \in D} {e^{{-||x-p||^2}\over{ 2\sigma^2}}}-\frac{\omega}{|S_n|} \sum_{p \in S_n}{e^{{-||x-p||^2}\over{2\sigma^2}}}
\]
where $\omega$ is a small positive number (typically in $(.1,.5)$).

This algorithm iteratively maximizes the function $F_n(x)$ by translating the points in $S_n$ in the direction of the gradient of $F_n(x).$  Let set $S_{n+1}$ be the points in $S_n$ translated in the direction of the gradient of $F_n(x)$ a distance proportional to $|\nabla F_n(p)|$ for each $p\in S_n$.  Namely, 
\[
S_{n+1}=\left\{p+c\frac{\nabla F_n(p)}{M} \mid p \in S_n\right\}
\]
where $\displaystyle M=\max_{p \in S_0}(|\nabla F_0(p)|)$, the maximum gradient observed in $F_0$, and $c$ is the maximum distance each point is allowed to translate at each step (typically $c$ is set to be proportional to the maximum inter-point distance in $D$).

\subsection{Motivation and Intuition}

The algorithm presented above is a modification of the mean-shift clustering algorithm presented by Fukunaga and Hostetler in \cite{fukunaga}.  Let 
\[
f(x)=\frac{1}{|D|}\sum_{p\in D} e^{{-||x-p||^2}\over{ 2\sigma^2}}
\]
Note that $f(x)$ is a kernel density estimator of the data set $D$.  The mean-shift clustering method iteratively translates the points in $D$ in the direction of the gradient of $ f(x)$.  The results of this clustering method is that the points converge to the maxima of the function $f(x)$.  In their paper, Fukunaga and Hostetler also note that in situations with only a small amount of noise, a small number of iterations of their method can be used as a data filtering method to recover the topology of the data (more iterations quickly sends the points to the maxima of $f(x)$).  Their algorithm fails to recover the topology in significant amounts of noise and quickly converges to the maxima of $f(x)$.

The intuition about how the algorithm in section~\ref{section:alg} recovers the underlying topology from a noisy data set is that the kernel density estimator $f(x)$ contains more information that just the maxima of the function.  We believe that significant topological information from the data set $D$ is found in regions of the domain where the density of the points is high, and thus the value of the kernel density estimator $f(x)$ is relatively high.  Moreover, we believe that the regions that have high density and the kernel density estimator $f(x)$ has small gradient are the most promising for the recovery of topological information.

\begin{figure}[h]
\centering
\fbox{\includegraphics[width=.4\textwidth]{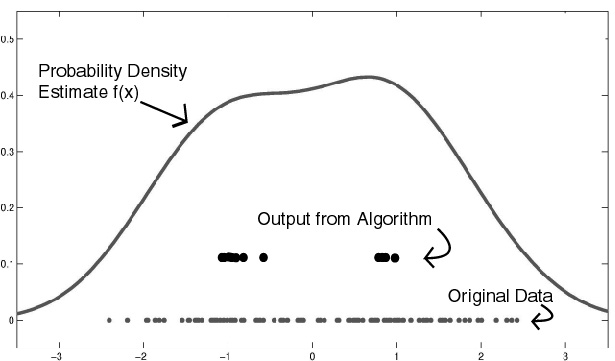}}
\caption{Intuition example.  Above is a one-dimensional data set $D$ sampled noisily from the points -1 and 1.  A kernel density estimate $f(x)$ is plotted above.  Notice that $f(x)$ has two regions that have high density but low gradient.  The output $S_{100}$ from the de-noising algorithm is shown slightly above the data set $D$.  Notice that the output is gathered over the regions of $f(x)$ that have high density and small gradient. }\label{fig:intuition}
\end{figure}

The idea is to fit another kernel density estimator, $g(x)$, to $f(x)$ in these regions.  The intuition is that low gradients on the kernel density function $f(x)$ often arise over the space $\mathbb{X}$ under Guassian-like noisy sampling of $\mathbb{X}$.  For example, if a space $\mathbb{X}$ has two components and is sampled with Guassian noise, the resulting kernel density estimator $f(x)$ will have a low gradient somewhere over each of the components of $\mathbb{X}$ even if the level of noise is very high.  The gradient need not be zero since the distance between the components may be small compared to the level of noise.  See figure~\ref{fig:intuition} for such an example.  By fitting the function $g(x)$ to the function $f(x)$ in these regions we find the positions of the points $S_n$ that could, under the estimated noise, generate similar density function estimator to the function $f(x)$ in these regions. 

Translating the points in $S_n$ in the direction of the function $F_n(x)$ allows for the recovery of the topology of the space $\mathbb{X}$.  While the first term of the function $F_n(x)$ (i.e. $f(x)$) serves to cluster the points in $S_n$ to the maxima $f(x)$, the second term in $F_n(x)$, $\omega g_n(x)$, serves to repel the points in $S_n$ away from each other.  When $\omega$ is a small positive number, the optimization of $F_n(x)$ is equivalent to fitting $g_n(x)$ to $f(x)$ in regions where the gradient of $f(x)$ is small and the value of $f(x)$ is relatively high (in other words, there is a significant density of points in $D$ in the region and the estimated density there is relatively flat).  This is because when the gradient of $f(x)$ is large, the second term is relatively small and the points in $S_n$ tend toward regions of high $f(x)$ values.  Within regions where $f(x)$ is relatively high and the gradient of $f(x)$ is low, the second term $\omega g_n(x)$ balances $f(x)$ and allows for the fitting of within these regions.

\section{Results with Synthetically-\\Generated Data}\label{section:3}
\subsection{Synthetically-Generated Noisy Circle}\label{section:cleancircle}
First we will apply the above algorithm to the data set $K$, the noisy sampling of the unit circle in $\mathbb{R}^2$, presented in section~\ref{sec:noisycircle}.  Recall that the data set $K$ shown in figure~\ref{fig:P7} is 1000 points sampled from the probability density function shown in figure~\ref{fig:Bump_7}.  Recall as was shown in section~\ref{sec:noisycircle}, density thresholding was unable to recover the correct first Betti number of the circle used to generate $K$ (see figures ~\ref{fig:barcodethresh} and ~\ref{fig:barcodethresh2}.)

\begin{figure}[h]
\centering
\fbox{\includegraphics[width=.25\textwidth]{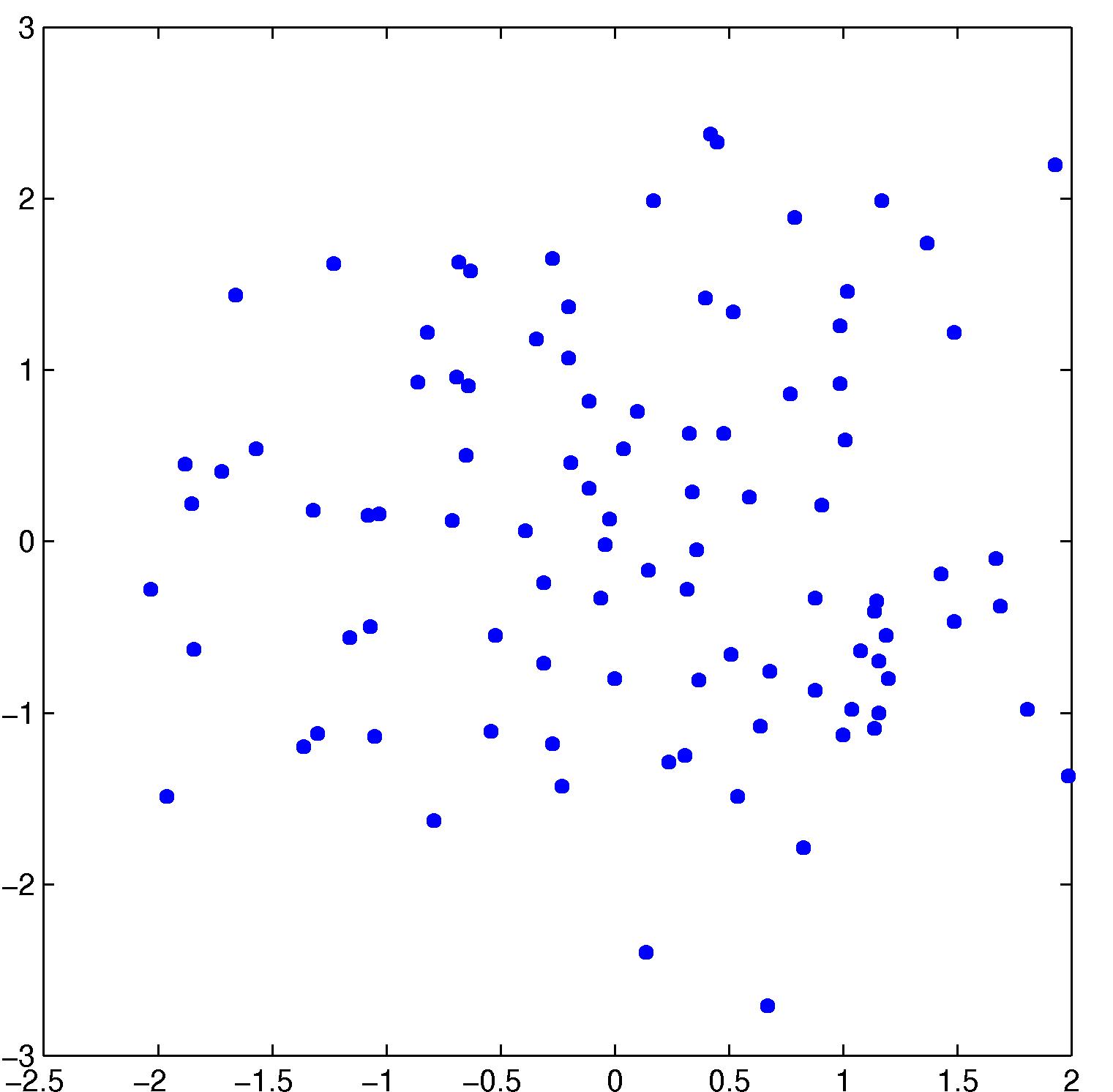}}
\caption{$S_0$ is 100 points selected randomly from $K$, the noisy circle data.}\label{fig:S_0}
\end{figure}

\begin{figure}[h]
\centering
\includegraphics[width=.45\textwidth]{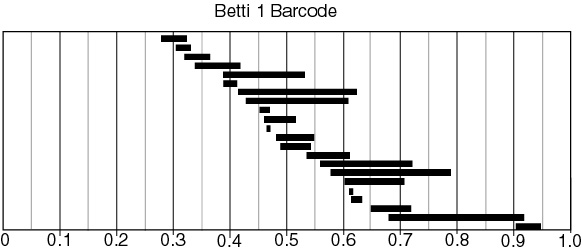}
\caption{Betti 1 persistence barcode of $S_{0}$}\label{fig:barcodeSP}
\end{figure}

\begin{figure}[h]
\centering
\includegraphics[width=.45\textwidth]{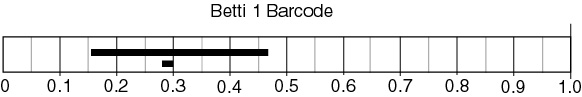}
\caption{Betti 1 persistence barcode of $S_{200}$, the output from the noisy circle data}\label{fig:barcodenps}
\end{figure}

To apply the above algorithm to $K$, we need a randomly selected subset of points $S_0$.  For this example, we chose $S_0$ to have 100 points randomly selected from $K$ (see figure~\ref{fig:S_0} for a plot of $S_0$ and figure~\ref{fig:barcodeSP} for the Betti 1 persistence barcode of $S_0$).  Recall that this algorithm has parameters for $\sigma$ the estimated standard deviation of the noise, $\omega$ the weighting factor for $F_n(x)$, and $c$, the maximal distance each point is allowed to translate per iteration.  Here we chose $\sigma = 0.6$, $\omega=0.1$ and $c=0.05$.  Figure~\ref{fig:S_{200}} shows $S_{200}$, the output  of this algorithm after 200 iterations.   Figure~\ref{fig:barcodenps} shows the Betti 1 persistence barcode generated from the Vietoris-Rips complex on $S_{200}$.  Notice the clean recovery of the circle in figure~\ref{fig:S_{200}} as well as the persistent first Betti number of one in figure~\ref{fig:barcodenps}.

Very similar persistence barcode results are obtained by letting $\sigma \in (0.4, 0.6)$ as well as when the number of points in the set $S_0$ ranges from $75$ to $200$.

\begin{figure}[h]
\centering
\fbox{\includegraphics[width=.25\textwidth]{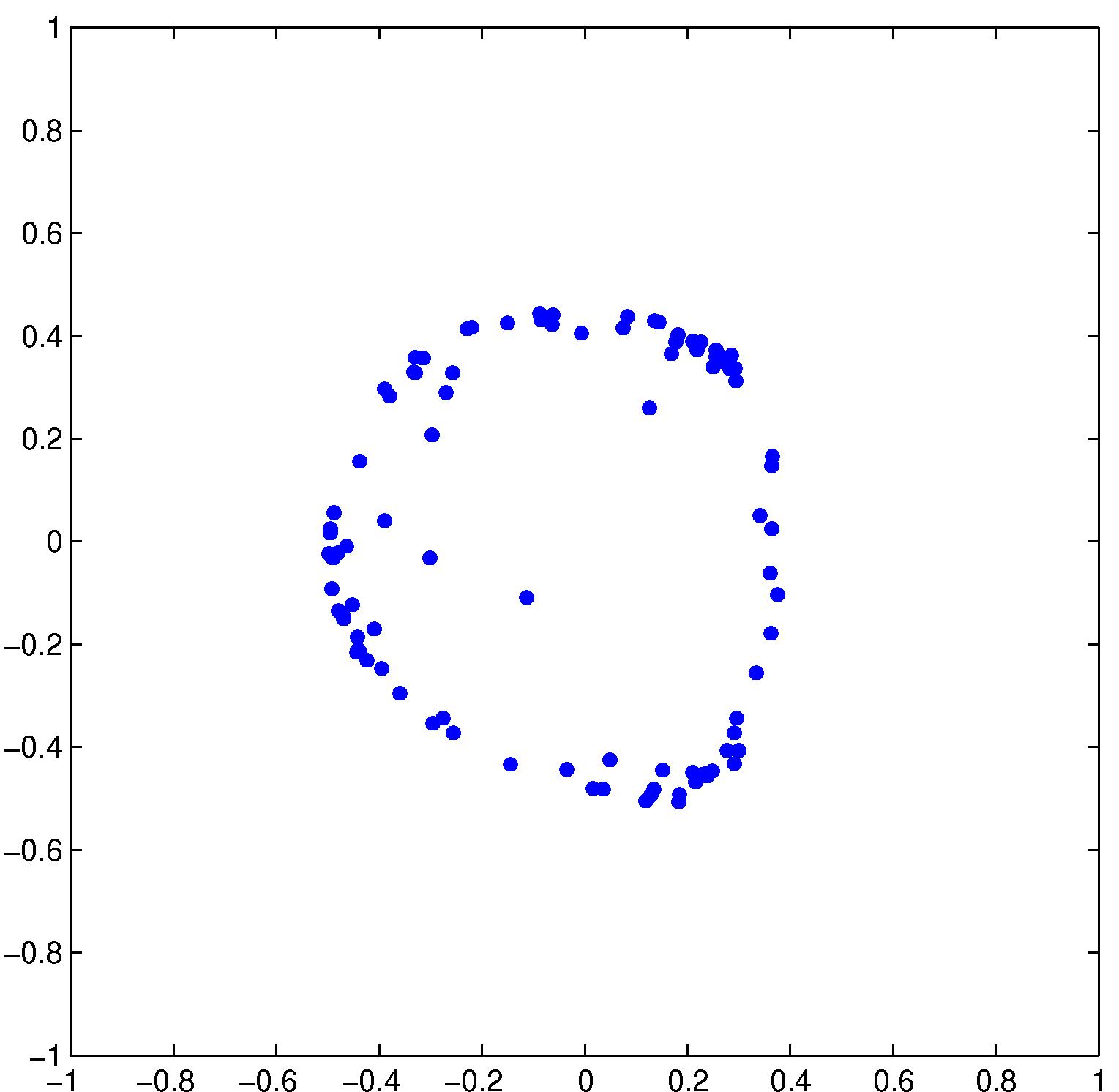}}
\caption{$S_{200}$, the output of the algorithm from section~\ref{section:alg} after 200 iterations applied $S_0 \in K$ with $\sigma=0.6$, $\omega=0.1$, and $c=0.05$}\label{fig:S_{200}}
\end{figure}

\subsection{Synthetically-Generated Noisy Sphere}

Next we apply the de-noising algorithm from section~\ref{section:alg} to a noisy sampling of the unit sphere in $\mathbb{R}^3$.  We generate this data set in a similar fashion to how the data set $K$ as generated in~\ref{section:bumpintro}.  Specifically, we sample the density function
\[
p(\vec{x})=\iint\limits_{\mathbb{R}^3} \, \delta_{S^2}(\vec{y}) e^{\frac{-||\vec{y}-\vec{x}||^2}{2(0.3)^2}}\,d\vec{y}
\]
where $\delta_{S^2}(\vec{y})$ is the delta function that has support being the unit sphere in $\mathbb{R}^3$ and the spherically-symmetric Gaussian function has standard deviation 0.3.  Let $L$ be the 1000 points sampled from the density function $p(\vec{x})$.

\begin{figure}[h]
\centering
\includegraphics[width=.45\textwidth]{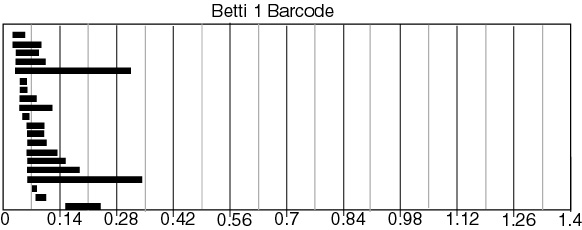}
\includegraphics[width=.45\textwidth]{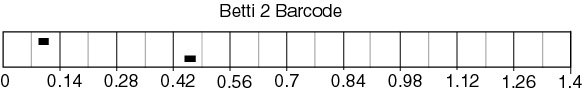}
\caption{Betti 1 and Betti 2 persistence barcode of noisy sphere.  This is generated from the $30\%$ densest points of $L$ using $15th$ nearest neighbor density estimation}\label{fig:barcodespherethresh}
\end{figure}

\begin{figure}[h]
\centering
\includegraphics[width=.45\textwidth]{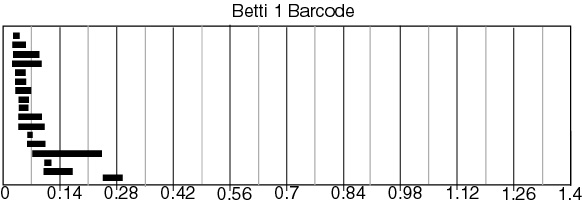}
\includegraphics[width=.45\textwidth]{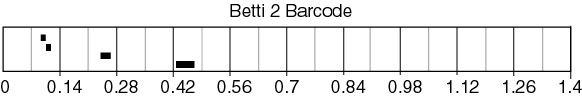}
\caption{Betti 1 and Betti 2 persistence barcode of noisy sphere.  This is generated from the $30\%$ densest points of $L$ using $30th$ nearest neighbor density estimation}\label{fig:barcodespherethresh2}
\end{figure}

Density thresholding on the data set $L$ fails to recover the underlying sphere. Figures~\ref{fig:barcodespherethresh} and~\ref{fig:barcodespherethresh2} are two examples of Betti 1 and Betti 2 persistent homology barcodes of density thresholding on the data set $L$.  The barcode in figure~\ref{fig:barcodespherethresh} was generated from a data set that contains the $30\%$ densest points of $L$ when the 15th nearest neighbor density estimator is used.  The barcode in figure~\ref{fig:barcodespherethresh2} was generated from a data set that contains the $30\%$ densest points of $S$ when the 30th nearest neighbor density estimator is used.  These barcodes were generated using lazy witness complexes on 100 randomly chosen landmarks (see \cite{witness} for a detailed discussion on lazy witness complexes).  Notice that neither barcode shows a second Betti number of one.

\begin{figure}[h]
\centering
\includegraphics[width=.45\textwidth]{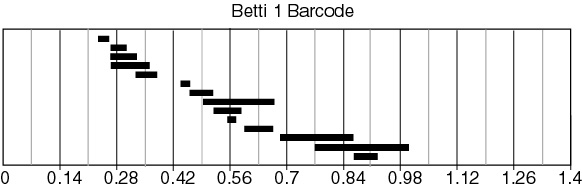}
\includegraphics[width=.45\textwidth]{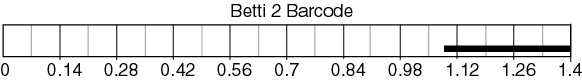}
\caption{Betti 1 and Betti 2 persistence barcode of $S_{200}$, the output from the noisy sphere $L$}\label{fig:barcodesphere}
\end{figure}

We can apply the algorithm from section~\ref{section:alg} to the data set $L$ as follows.  Let $S_0$ be a randomly chosen 100 point subset of the data set $L$.  Run the algorithm for 200 iterations with $\sigma=0.35$ and $c=0.05$.  The resulting data set $S_{200}$ was used to create the Vietoris-Rips complex that generated the barcodes in figure~\ref{fig:barcodesphere}.  Notice the persistent second Betti number is clearly one and there is no persistent first Betti number.

\subsection{Synthetically-Generated Noisy Point}

\begin{figure}[h]
\centering
\fbox{\includegraphics[width=.25\textwidth]{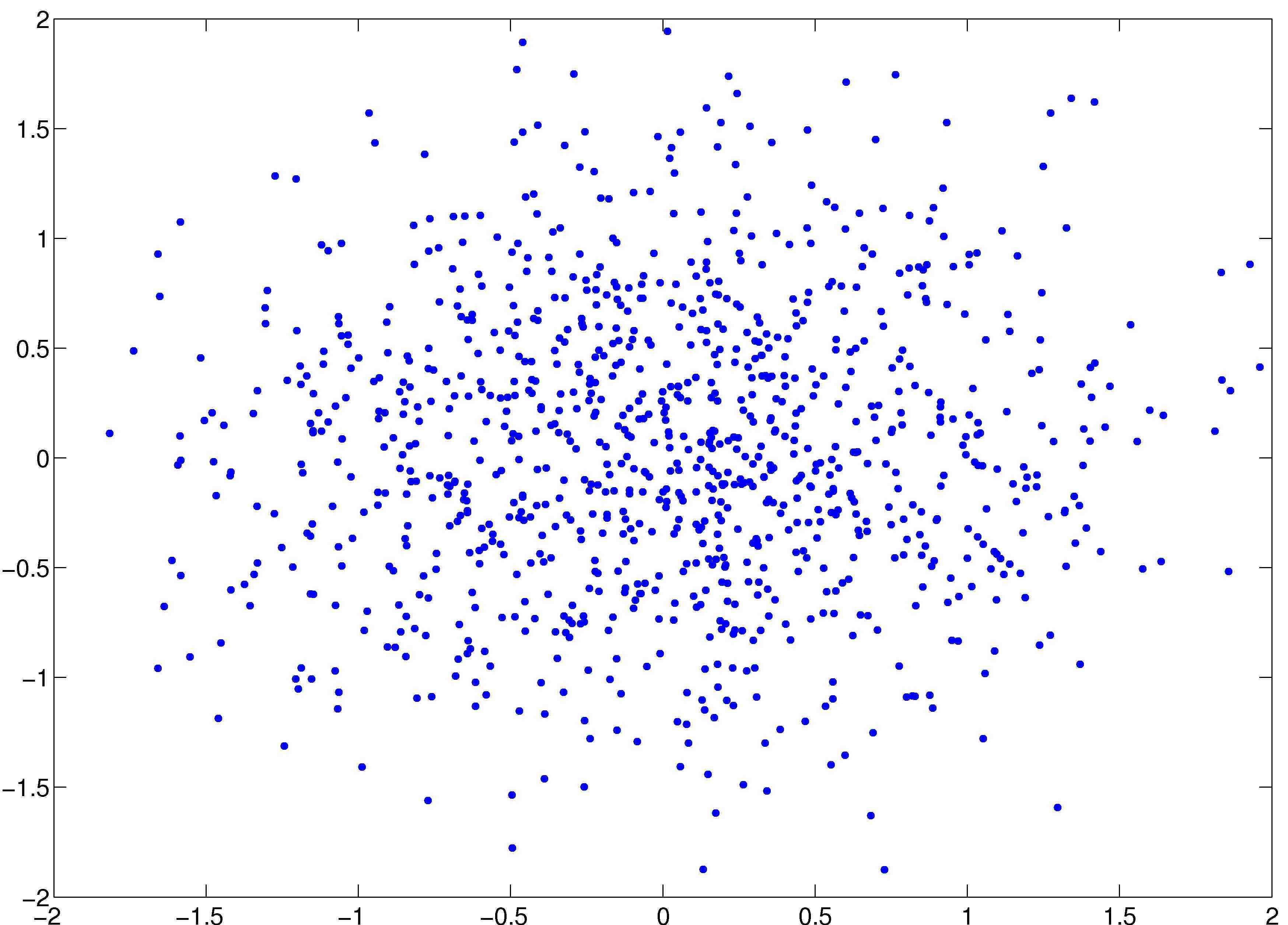}}
\fbox{\includegraphics[width=.25\textwidth]{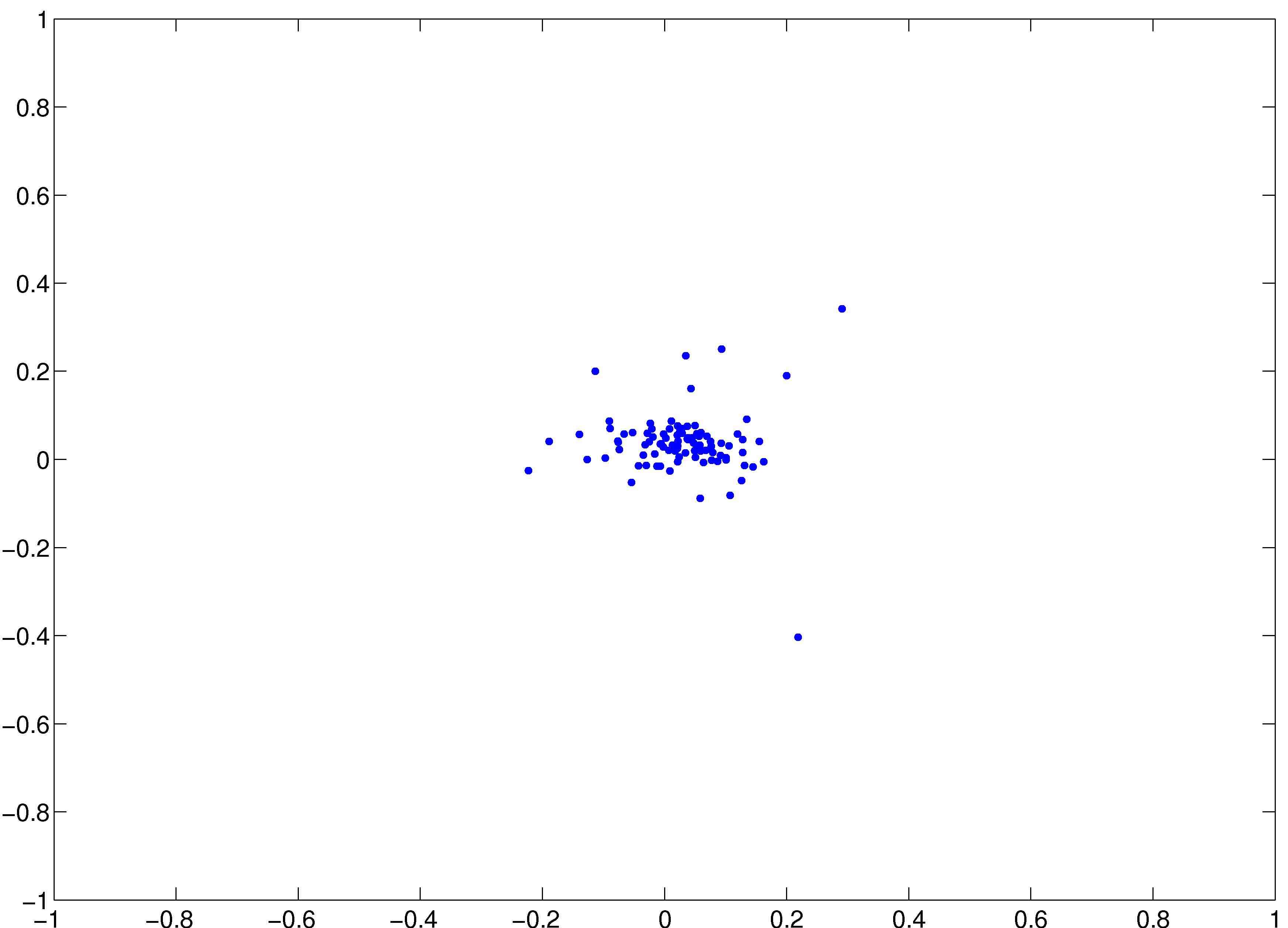}}
\caption{A noisy sampling $P$ of a point in $\mathbb{R}^2$ followed by the output $S_{50}$ where $\sigma = 0.6$ and $\omega = 0.1$.}\label{fig:point}
\end{figure}

\begin{figure}[h]
\centering
\includegraphics[width=.45\textwidth]{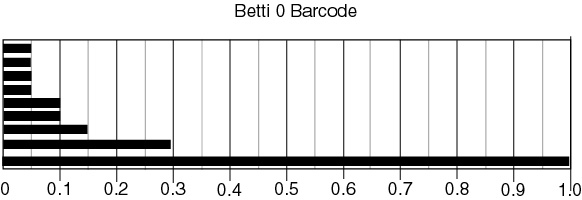}
\includegraphics[width=.45\textwidth]{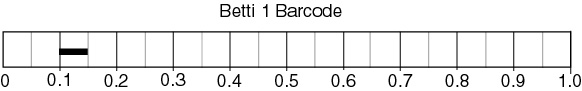}
\caption{Betti 0 and Betti 1 persistence barcode of $S_{50}$, the output from the noisy point $P$ in figure~\ref{fig:point}}\label{fig:pointbarcode}
\end{figure}

For the sake of completeness, we have included the results of our algorithm applied to a noisy sampling of a point in $\mathbb{R}^2$.  All parameters, except the number of iterations, were chosen to be those from the noisy sampling of the circle presented above in section~\ref{section:cleancircle}.  The results below are produced with only 50 iterations.  After 50 iterations, the algorithm correctly clustered all the points to a point.  For the sake of presentation, we show the results at fewer iterations.  The original data set and the resulting output is shown in figure~\ref{fig:point}.  The Betti 0 and Betti 1 barcode outputs for the output from our algorithm are in figure~\ref{fig:pointbarcode}. 

\section{Results from Image Processing Data}\label{section:mumford}

Natural image statistics have attracted much interest in recent years.  In \cite{witness}, \cite{localbehavior}, and \cite{tigran}, persistent homology was used to understand the local behavior of natural images.  Instead of viewing the image as a whole, they analyzed the structure of high-contrast pixel patches.  Each of the papers above analyzed the data set $M$ that is a collection of $4\cdot 10^6$ `3 by 3' high contrast pixel patches obtained from the collection of still images gathered by J. H. van Hateren and A. van der Schaaf \cite{hateren}.  $M$ is a subset of a larger set $\tilde{M}$ with $8 \cdot 10^6$ images provided by K. Pedersen.

The space $\tilde{M}$ is obtained by applying the following procedure to a subset of images from a still image collection  (See \cite{mumford} for more details). 

\begin{enumerate}
\item Select an image from the still image collection.
\item Extract at random 5000 `3 by 3' patches from the image.  Regard each patch as a vector in 9-dimensional space.
\item Next, for each patch do the following:
\begin{enumerate}
\item Compute the logarithm of intensity at each pixel.
\item Subtract an average of all coordinates from each coordinate.  This produces a new 9-vector.
\item For this vector of logarithms compute the contrast or ``$D$-norm'' of the vector.  The $D$-norm of a vector $x$ is defined as $\sqrt{x^TDx}$, where $D$ is a certain positive definite symmetric $9 \times 9$ matrix. 
\item Keep this patch if its $D$-norm is among the top 20 percent of all patches taken from the image.
\end{enumerate}
\item Normalize each of the selected vectors by dividing by their respective $D$-norms.  This places them on the surface of a 7-dimensional ellipsoid.
\item Perform a change of coordinates so that the resulting set lies on the actual 7-dimensional sphere in $\mathbb{R}^8.$
\end{enumerate}

\subsection{Previous Persistence Results with Thresholding}

\begin{figure}[h]
\centering
\includegraphics[width=.45\textwidth]{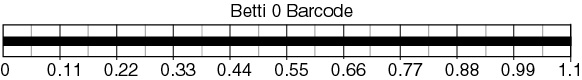}
\includegraphics[width=.45\textwidth]{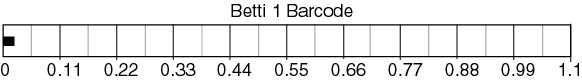}
\caption{Betti 0 and 1 persistence barcodes of a subset of $M$ of size $10^5$.}\label{fig:mumford}
\end{figure}

Because the space of patches $M$ is distributed throughout the unit disk in $\mathbf{R}^8$, not just in high-density spaces, all previous topological analysis of this space has required density thresholding.  Figure~\ref{fig:mumford}, a Betti 1 persistence barcode on a subset of $M$ of size $10^4$, illustrates the inability of persistence homology to recover any topological structure without some kind of filtering or de-noising.  One example of this is in \cite{witness}, where persistence homology was used to analyze dense subspaces within the space of patches.  Working on a randomly chosen subset of $M$ containing $5\cdot 10^4$ points, the authors of \cite{witness} showed that the $30\%$ densest points in this subset, using the 300th nearest neighbor density estimator, were gathered around an annulus.  They showed that this dense annulus can be characterized by the angle of the edge separating a dark region of the pixel patch from a light region.

\begin{figure}[h]
\centering
\fbox{\includegraphics[width=.25\textwidth]{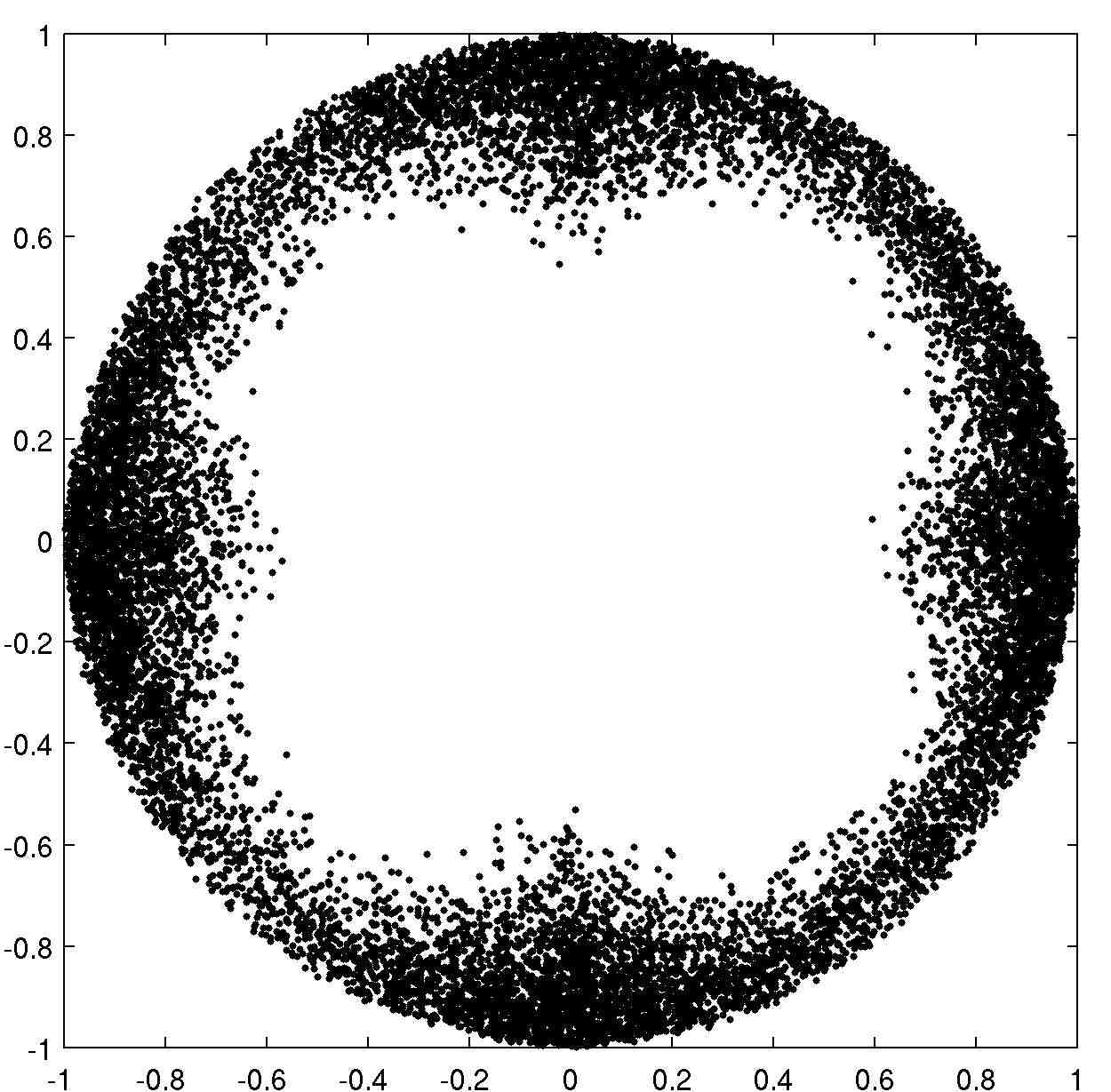}}
\caption{First two coordinates of 30$\%$ densest points of a $50,000$ point subset of $M$ using 300th nearest neighbor density estimator}\label{fig:Mumfordtop30}
\end{figure}

\begin{figure}[h]
\centering
\includegraphics[width=.45\textwidth]{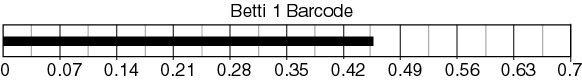}
\caption{Betti 1 persistence barcode of $30\%$ densest points of a $50,000$ point subset of $M$ using 300th nearest neighbor density estimator}\label{fig:BarcodeMumfordtop30}
\end{figure}

This annulus can clearly be seen in the plot of the first two dimensions of a similarly generated dense subset of $M$ in figure~\ref{fig:Mumfordtop30}.  Figure~\ref{fig:BarcodeMumfordtop30} is the Betti 1 persistence barcode for this data set.  This barcode is generated using a lazy witness complex built on 100 randomly chosen landmarks.

\subsection{Results of De-Noising Algorithm}

We can obtain similar results from the data set $M$ by using the algorithm from section~\ref{section:alg} instead of thresholding.  Let $\bar{M}$ be a randomly chosen subset of $M$ that contains $1\cdot 10^5$ points and let $S_0$ be a randomly chosen subset of $\bar{M}$ that contains 500 points.  The algorithm was applied to $\bar{M}$ and $S_0$ with three different values of $\sigma$.  Namely, when $\sigma = 0.4, 0.5$, and $ 0.6$.  Each instance of the algorithm was run for 100 iterations, $c=0.05$ and $\omega=0.1$.

\begin{figure}[h]
\centering
\includegraphics[width=.45\textwidth]{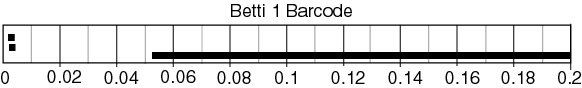}
\caption{Betti 1 persistence barcodes of $S_{100}$ applied to $\bar{M}$ using $\sigma = 0.4$}\label{fig:barcodeMumford66}
\end{figure}

\begin{figure}[h]
\centering
\includegraphics[width=.45\textwidth]{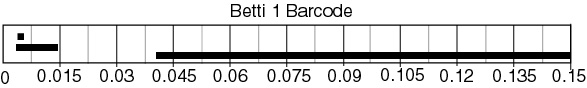}
\caption{Persistence barcodes of $S_{100}$ applied to $\bar{M}$ using $\sigma = 0.5$}\label{fig:barcodeMumford77}
\end{figure}

\begin{figure}[h]
\centering
\includegraphics[width=.45\textwidth]{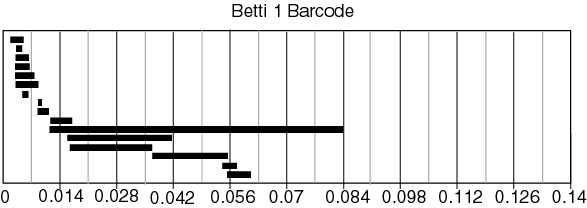}
\caption{Betti 1 persistence barcodes of $S_{100}$ applied to $\bar{M}$ using $\sigma = 0.6$}\label{fig:barcodeMumford88}
\end{figure}

\begin{figure}[h]
\centering
\fbox{\includegraphics[width=.25\textwidth]{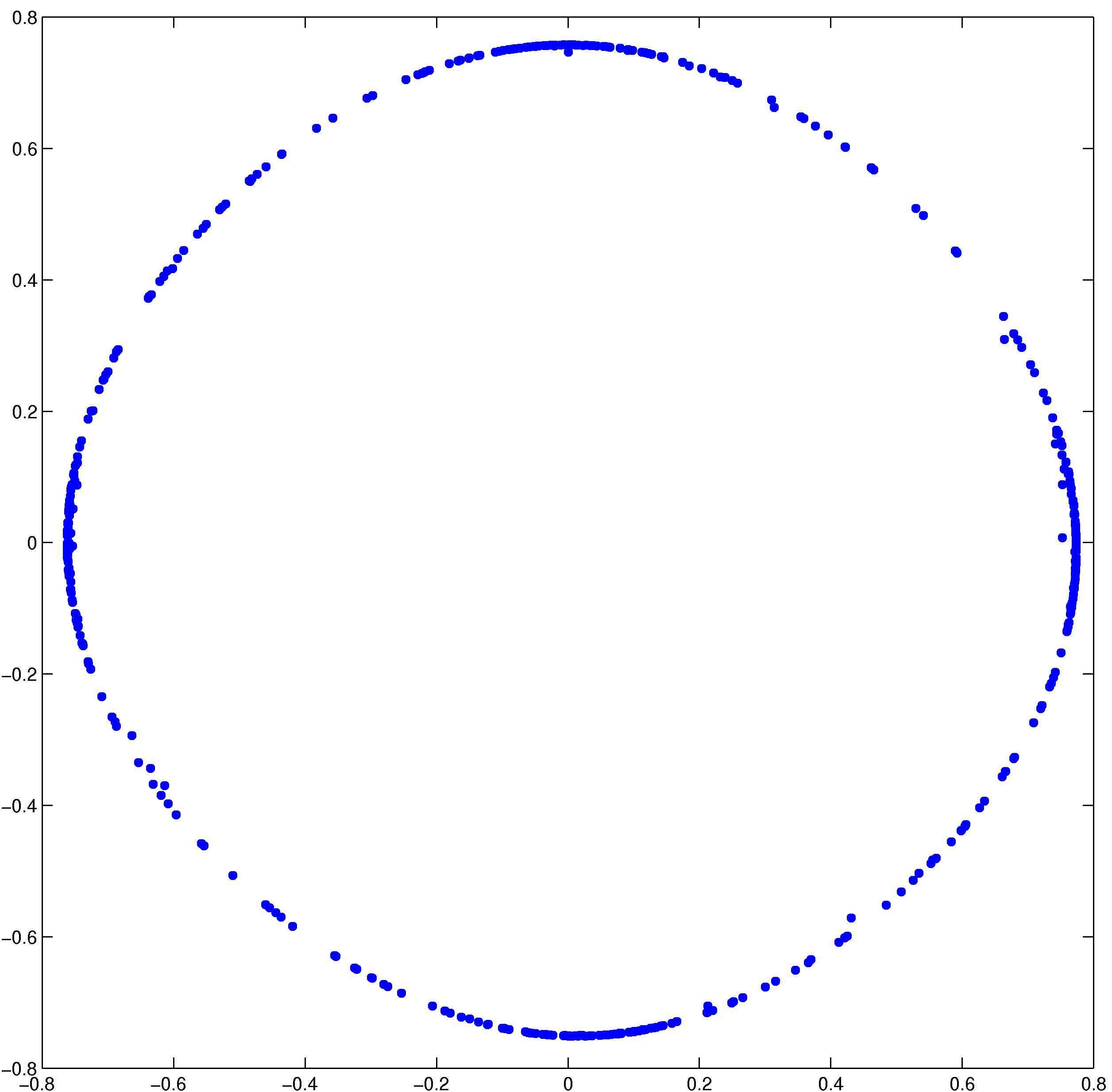}}
\caption{First two coordinates of  $S_{100}$ applied to $M$ using $\sigma = 0.4$}\label{fig:Mumford66plot}
\end{figure}

\begin{figure}[h]
\centering
\fbox{\includegraphics[width=.25\textwidth]{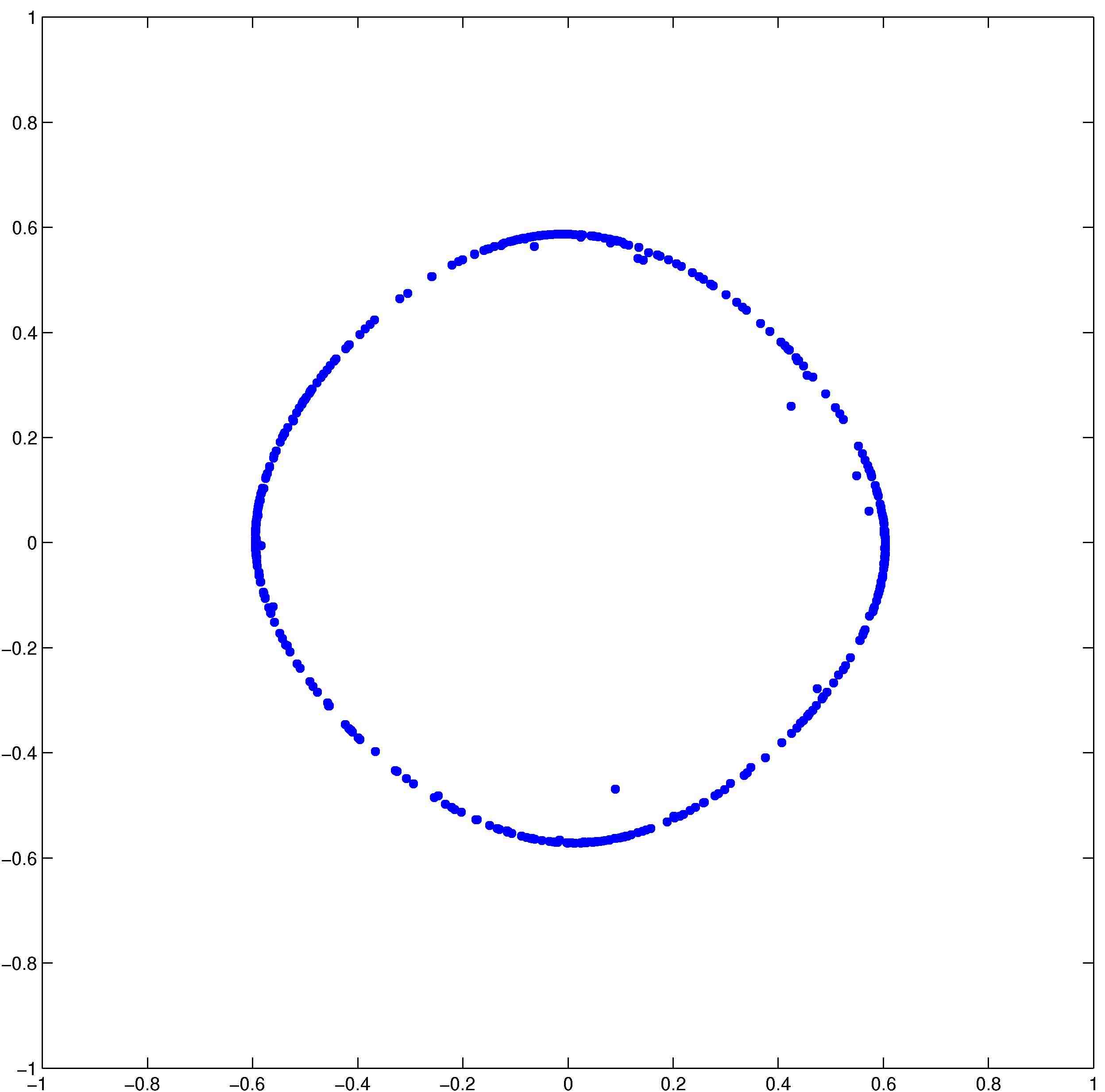}}
\caption{First two coordinates of  $S_{100}$ applied to $M$ using $\sigma = 0.5$}\label{fig:Mumford77plot}
\end{figure}

\begin{figure}[h]
\centering
\fbox{\includegraphics[width=.25\textwidth]{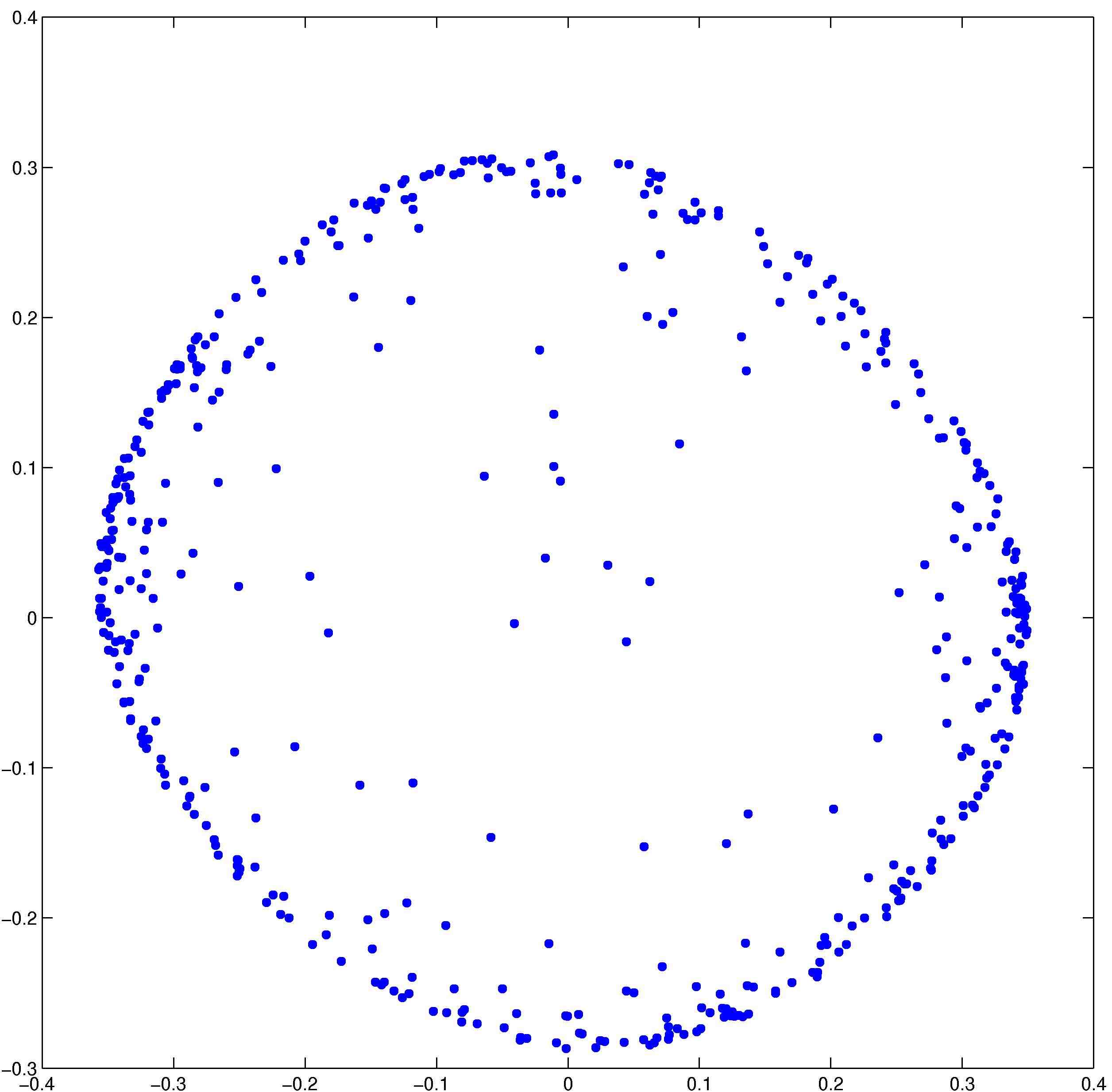}}
\caption{First two coordinates of  $S_{100}$ applied to $M$ using $\sigma = 0.6$}\label{fig:Mumford88plot}
\end{figure}

Figures~\ref{fig:barcodeMumford66},~\ref{fig:barcodeMumford77}, and~\ref{fig:barcodeMumford88} show the Betti 1 persistence barcodes generated on the de-noised data sets $S_{100}^{\sigma=0.4}, S_{100}^{\sigma=0.5},$ and $S_{100}^{\sigma=0.6}$ respectively.  Each barcode is generated using a lazy witness complex built on 100 randomly chosen landmarks.  For reference, the plot of the first two coordinates of the dense subset shown in  figure~\ref{fig:Mumfordtop30}, figures~\ref{fig:Mumford66plot},~\ref{fig:Mumford77plot}, and~\ref{fig:Mumford88plot} are the respective plots of the first two coordinates of $S_{100}^{\sigma=0.4},$ $ S_{100}^{\sigma=0.5},$ and $S_{100}^{\sigma=0.6}.$  Notice the clear presence of a persistent first Betti number of one in each denoised data set.


\section{Results from Range Image Data}\label{section:mumford}

An optical image has a grayscale value at each pixel, whereas a range image pixel contains a distance: the distance between the laser scanner and teh nearest object in the correct direction.  As before with the natural image data, we can think of an $n \times m$ pixel patch as a vector in $\mathbb{R}^{n \times m}$ and a set of patches as a set of points in $\mathbb{R}^{n \times m}$.  In \cite{hadams}, persistent homology was used to understand the local behavior of range image patches.  As was done in the natural image work, the data studied was high-contrast range pixel patches.  Previous work done on range image patches in \cite{mumford} found that $3 \times 3$ range patches simply broke into clusters without any obvious simple geometry.  In \cite{hadams}, the study was extended to larger $5 \times 5$ and $7 \times 7$ range patches.  

These patches were normalized in much of the same was as the natural image data described in section~\ref{section:mumford}.  The resulting data set contains 50,000 high-contrast, normalized range image patches.

\begin{figure}[h]
\centering
\includegraphics[width=.45\textwidth]{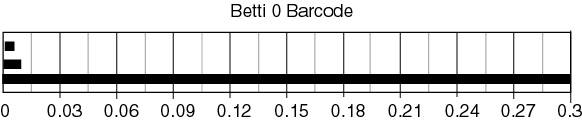}
\includegraphics[width=.45\textwidth]{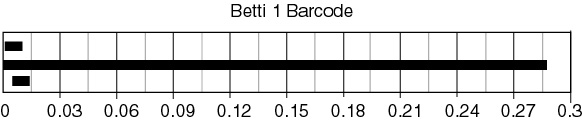}
\caption{Betti 0 and 1 persistence barcodes the $30\%$ densest points in the 7 $\times$ 7 range image patch data.  Here the 300th nearest neighbor density estimate was used.}\label{fig:hadams}
\end{figure}

Similar to the results on the natural image data, the authors of \cite{hadams} showed that the $30\%$ densest points in this subset, using the 300th nearest neighbor density estimator, were gathered around an annulus.  They showed that this dense annulus can be characterized by the angle of the linear gradients between a dark region of the pixel patch to a light region.  Figure~\ref{fig:hadams} contains the Betti 0 and Betti 1 barcodes for $30\%$ densest points in the $7 \times 7$ image range patches where 300th nearest neighbor density estimator was used.

\begin{figure}[h]
\centering
\includegraphics[width=.45\textwidth]{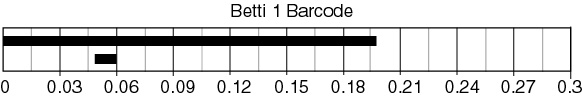}
\includegraphics[width=.45\textwidth]{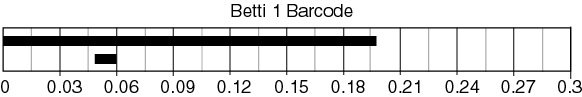}
\caption{Betti 0 and 1 persistence barcode of $S_{100}$ applied to the 7 $\times$ 7 range image data using $\sigma = 0.5$ and $\omega = 0.1$.}\label{fig:rangeimage}
\end{figure}

Figure~\ref{fig:rangeimage} contains the results of the algorithm from section~\ref{section:alg} applied to a randomly chosen 5000 point subset of the $7 \times 7$ range image patches.  These results were generated using 100 iterations of the algorithm with $\sigma = 0.5$ and $\omega = 0.1$.

\section{Discussion and Future Directions}
The algorithm presented in section~\ref{section:alg} uses spherically-symmetric Gaussian functions to estimate the density of the generating probability density functions used $F_n(x)$.  It is likely that similar spherically-symmetric kernels such as the Epanechnikov kernel (see page 76 of \cite{silverman} for more details) would produce similar results.  The use of a spherically-symmetric kernel implies that the data and noise are symmetric.  In certain circumstances, it may be appropriate to pre-scale the data to avoid extreme differences in the spread in the various coordinate directions \cite{silverman}.\\

A significant benefit of this method is that it performs quickly even when the ambient dimension is high.  Specifically, the runtime analysis of the algorithm is $d^2NMn$ where $d$ is the dimension of the ambient Euclidean space which contains data set, $N$ is the number of data points in the data set $D$, $M$ is the number of points in the set $S_0$, and $n$ is the number of iteration of the algorithm.  Moreover, the implementation of this algorithm is very similar to that of the mean-shift algorithm and hence it should not be difficult to implement in a parallel programming paradigm.\\

\subsection{Stopping Criterion, Ill-posed Problem, and Zig-Zag Persistence}

As $n$ increases, the data sets $S_n$ tend towards the same topology as $\mathbb{X}$ while decreasing the amount of noise.  Since the convergence of this algorithm depends on the choice of $S_0$, it is unclear at which $n$ to stop this algorithm.  If the researcher has prior knowledge about the underlying space $\mathbb{X}$ such as its dimension, then you can stop when $S_n$ has the same estimated dimension.  In the future we plan to use a new method known as zig-zag persistence (see \cite{zigzag}) to determine the stopping criterion in the instance when there is no prior knowledge about the space $\mathbb{X}.$\\

Note also that this algorithm is an ill-posed problem in that solutions obtained from applying this technique to a different subset $S'_0 \subset D$ will produce different output.  While we expect most of the output of this algorithm to retain the correct topology of $D$, it is possible that some output will not recover the correct topology.\\

Zig-zag persistence, which is a new method for studying persistence of topological features across a family of point-cloud data sets regardless of the direction of the arrows between these sets, should allow us to obtain a measure of confidence in our results as well as to determine an appropriate stopping criterion.  Let $\{ S_0^i\}_{i=1}^m$ be a collection of $m$ randomly chosen subsets of $D$.  Run the algorithm above on each set $S_0^i$ a fixed number of times, say $n$ (for instance, $n=\frac{d}{2c}$ where $d$ is the largest distance between any two points in the original data set $D$).  We then obtain the following sequence of data sets:

\[
S_n^1 \hookrightarrow S_n^1 \cup S_n^2 \hookleftarrow S_n^2 \hookrightarrow S_n^2\cup S_n^3 \hookleftarrow ... \hookleftarrow S_n^m
\]

In the future we plan to use zig-zag persistence to correlate topological features across these different instances of this de-noising algorithm to give us an accurate picture of the topology of the data set $D$.  In other words, if the majority of the $S_n^i$ show a second Betti number of 1, zig-zag persistence should tell whether persistence homology is detecting the same class generating the 2-cycle in each space or whether there are many different 2-cycles each observed in different spaces.  Moreover, even if not all the $S_n$'s have converged to the point that persistent homology techniques can recover the topological features of $D$, zig-zag persistence should allow us to detect features present in any of the $S_n$'s as well as knowledge about how these features persist through the various $S_n$'s.\\

\bibliographystyle{plain}
\bibliography{TopologicalDeNoisingPaper2}
\end{document}